\documentstyle[epsf,psfig]{mn}

\newif\ifAMStwofonts

\ifoldfss
  \ifCUPmtlplainloaded \else
    \NewTextAlphabet{textbfit} {cmbxti10} {}
    \NewTextAlphabet{textbfss} {cmssbx10} {}
    \NewMathAlphabet{mathbfit} {cmbxti10} {} 
    \NewMathAlphabet{mathbfss} {cmssbx10} {} 
  \fi
  \ifAMStwofonts
    \ifCUPmtlplainloaded \else
      \NewSymbolFont{upmath} {eurm10}
      \NewSymbolFont{AMSa} {msam10}
      \NewMathSymbol{\upi}     {0}{upmath}{19}
      \NewMathSymbol{\umu}     {0}{upmath}{16}
      \NewMathSymbol{\upartial}{0}{upmath}{40}
      \NewMathSymbol{\leqslant}{3}{AMSa}{36}
      \NewMathSymbol{\geqslant}{3}{AMSa}{3E}

    \fi
  \fi
\fi 

\ifnfssone
  \newmathalphabet{\mathit}
  \addtoversion{normal}{\mathit}{cmr}{m}{it}
  \addtoversion{bold}{\mathit}{cmr}{bx}{it}
  \newmathalphabet{\mathbfit} 
  \addtoversion{normal}{\mathbfit}{cmr}{bx}{it}
  \addtoversion{bold}{\mathbfit}{cmr}{bx}{it}
  \newmathalphabet{\mathbfss} 
  \addtoversion{normal}{\mathbfss}{cmss}{bx}{n}
  \addtoversion{bold}{\mathbfss}{cmss}{bx}{n}
  \ifAMStwofonts
    \ifCUPmtlplainloaded \else
      \UseAMStwoboldmath
      \makeatletter
      \new@mathgroup\upmath@group
      \define@mathgroup\mv@normal\upmath@group{eur}{m}{n}
      \define@mathgroup\mv@bold\upmath@group{eur}{b}{n}
      \edef\UPM{\hexnumber\upmath@group}
      \new@mathgroup\amsa@group
      \define@mathgroup\mv@normal\amsa@group{msa}{m}{n}
      \define@mathgroup\mv@bold\amsa@group{msa}{m}{n}
      \edef\AMSa{\hexnumber\amsa@group}
      \makeatother
      \mathchardef\upi="0\UPM19
      \mathchardef\umu="0\UPM16
      \mathchardef\upartial="0\UPM40
      \mathchardef\leqslant="3\AMSa36
      \mathchardef\geqslant="3\AMSa3E
    \fi
  \fi
\fi 

\ifnfsstwo
  \DeclareMathAlphabet{\mathbfit}{OT1}{cmr}{bx}{it}
  \SetMathAlphabet\mathbfit{bold}{OT1}{cmr}{bx}{it}
  \DeclareMathAlphabet{\mathbfss}{OT1}{cmss}{bx}{n}
  \SetMathAlphabet\mathbfss{bold}{OT1}{cmss}{bx}{n}
  \ifAMStwofonts
    \ifCUPmtlplainloaded \else
      \DeclareSymbolFont{UPM}{U}{eur}{m}{n}
      \SetSymbolFont{UPM}{bold}{U}{eur}{b}{n}
      \DeclareSymbolFont{AMSa}{U}{msa}{m}{n}
      \DeclareMathSymbol{\upi}{0}{UPM}{"19}
      \DeclareMathSymbol{\umu}{0}{UPM}{"16}
      \DeclareMathSymbol{\upartial}{0}{UPM}{"40}
      \DeclareMathSymbol{\leqslant}{3}{AMSa}{"36}
      \DeclareMathSymbol{\geqslant}{3}{AMSa}{"3E}
    \fi
  \fi
\fi 

\ifCUPmtlplainloaded \else
  \ifAMStwofonts \else 
    \def\upi{\pi}
    \def\umu{\mu}
    \def\upartial{\partial}
  \fi
\fi

\title{Optical counterparts to four X-ray sources in the Small Magellanic
 Cloud.}
\author[W. R. T. Edge]
       {W. R. T. Edge, M. J. Coe \\
        Department of Physics and Astronomy, Southampton University, SO17 1BJ}
\date{Accepted 2002.
      Received 2002;
      in original form 2002}

\pagerange{\pageref{firstpage}--\pageref{lastpage}} \pubyear{2002}

\begin{document}

\maketitle

\label{firstpage}

\begin{abstract}

We report on the spectroscopic and photometric analysis of possible optical
counterparts to four X-ray sources in the Small Magellanic Cloud: AX
J0049.4-7323, AX J0057.4-7325, RX J0058.2-7321 and RX J0101.1-7206. In the
case of the last two, we suggest that the presence of strong H$\alpha $
emission from previously proposed optical candidates is definite
confirmation that these are Be stars. Similarly, we detected strong
 H$\alpha $ emission from the optical source identified with RX J0049.7-7323
 within the error circle for AX J0049.4-7323 and we conclude that these are one
and the same object. We were unable to detect any H$\alpha $ emission from any
of the candidates for AX J0057.4-7325 and the associated photometric analysis
was also inconclusive.

\end{abstract}


\section{INTRODUCTION}
\label{sec:introductionntificat}

\subsection{The Magellanic Clouds}
\label{subsec:mylabel1}

The Magellanic Clouds are a pair of satellite galaxies which are
gravitationally bound to our own but which have characteristics
differing significantly from each other, and from the Milky Way.
Both clouds are relatively low in metal abundances reflecting
their separate and distinct evolutionary development (Matteucci et
al 2002). These differences in chemical composition are likely to
be reflected in the properties of different stellar populations.
The Small Magellanic cloud (SMC) is located at a distance of about
65 kpc and centred on a position of R.A. 1hr Dec. -73$^{o}$. It is
therefore close enough to be observed with modest ground based
telescopes but at the same time, because of its structural and
chemical differences, it provides an opportunity to study the way
in which these factors may influence the evolution of other
galaxies.

Physical characteristics such as mass distribution, orbital period
and spectral type are measurable parameters which are likely to
yield useful information in the study of differences between the
X-ray binary populations of the Magellanic Clouds and our own
Galaxy. To find out whether these differ significantly, it is
important to identify the optical counterparts of those X-ray
sources which remain unidentified and to study as many systems as
possible so as to increase the size of the sample studied to a
statistically significant number (Coe {\&} Orosz 2000).

\subsection{High Mass X-ray Binaries in the SMC}
\label{subsec:mylabel2}

Intensive X-ray satellite observations of the SMC have revealed
that it contains an unexpectedly large number of High Mass X-ray
Binaries (HMXB). These are binary pairs consisting of an early
type star in the spectral range O8-B2 and an accreting neutron
star, or, more rarely, a black hole. If the compact object in the
binary is a neutron star, it is likely to have a strong magnetic
field so that matter that has been gravitationally captured, it is
trapped by the field lines. The gravitational energy of the
infalling material is then released in the form of X-rays that
will be observed as pulses as the star rotates. The donor star in
the binary pair is likely to be visible in the infrared, optical
or ultraviolet, whereas the neutron star can generally only be
seen in the X-ray range (Negueruela 1998).

At the time of writing, 34 known or probable sources of this type
have been identified in the SMC and they continue to be discovered
at a rate of 1-2 per year, although only a small fraction of these
are active at any one time because of their transient nature. All
X-ray binaries so far discovered in the SMC are HMXBs (Coe 2000).
Observations of these appear to show marked differences by
comparison with the Galactic population. The X-ray luminosity
distribution of the Magellanic Clouds sources appears to be
shifted to higher luminosities and there also seems to be a higher
incidence of sources suspected of containing black holes (Stevens
et al, 1999).

\subsection{Be/X-ray Binaries}
\label{subsec:mylabel4}

HMXBs are divided into two broad classes on the basis of the
spectral type of the optical counterpart. The first category
includes the supergiant X-ray binaries (SXBs) in which the compact
object accretes from an OB supergiant leading to persistent X-ray
emission. The other group comprises the Be/X-ray binaries in which
a neutron star orbits an OB star surrounded by a circumstellar
disc of variable size and density (Negueruela and Coe 2002). In
general, these two subgroups are identified with persistent and
transient X-ray sources respectively.

Most HMXBs belong to the Be class, and all except one in the SMC
are in this category. The optical companion stars are early-type
luminosity class III-V, typically of 10 to 20 solar masses that at
some time have shown emission in the Balmer series lines. The
systems as a whole exhibit significant excess flux at long (IR and
radio) wavelengths, referred to as the infrared excess. These
characteristic signatures as well as strong H$\alpha $ line
emission are attributed to the presence of circumstellar material
in a disk like configuration (Coe 2000, Okazaki and Negueruela
2001).

The mechanisms which give rise to the disk are not well
understood, although fast rotation is likely to be an important
factor, and it is possible that non-radial pulsation and magnetic
loops may also play a part. The disk is thought to consist of
relatively cool material, which is intercepted periodically by the
compact object in an eccentric orbit, leading to regular X-ray
outbursts. Another possibility is that the Be star undergoes a
sudden ejection of matter (Negueruela 1998). A common feature of
the disk is the so-called global one-armed oscillation: an
asymmetric distribution of gas in the disk which is detected as
V/R asymmetry in the H$\alpha $ line profile (Okazaki 1993). This
is a characteristic feature of many Be type spectra and may be a
consequence of variations in orbital velocity and gas density due
to Keplerian motion in an eccentric gas ring. The V/R peak
intensity ratios often show cyclic variations ranging from a few
years to several decades and these are ascribed to slow apsidal
motion of the gas ring in the gravitational field of the central
star (Dachs et al 1992).

Be/X-ray binaries can present differing states of X-ray activity
varying from persistent low or non-detectable luminosity to short
outbursts, the latter usually coinciding with the periastron of
the neutron star. The low-luminosity persistent emission is due to
accretion of low density material which could be wind driven but
is more likely to be equatorial outflow beyond the regions in
which rotation dominates. Systems with small orbits will tend to
accrete from dense regions of the disk over a range of orbital
phases and give rise to very high luminosity outbursts, although
these may be modulated by the presence of a density wave in the
disk. Systems with wider orbits will tend to accrete from less
dense regions and hence show smaller outbursts, often only at
periastron passage (Negueruela 1998).

\section{Optical and IR Data}

Spectroscopic observations in H$\alpha $ of possible optical
counterparts to approximately 50 X-ray sources were made between
2001 November 6-12, using the SAAO 1.9m telescope. A 1200 lines
mm$^{ - 1}$ reflection grating blazed at 6800{\AA} was used with
the SITe CCD which is effectively 266 x 1798 pixels in size,
creating a wavelength coverage of  6160{\AA} to 6980{\AA}. The
intrinsic resolution in  this mode was 0.42{\AA}/pixel.

Optical photometric observations were taken from the SAAO 1.0m telescope during
 January 2000 and, in the case of RX J0101.0-7206, were supplemented by
 observations made earlier, using the same system, on
1996 October 2. The data were collected using the Tek8 CCD giving
a field of $ \sim $6x6 arc minutes and a pixel scale of 0.6
arcsec/pixel. Observations were made through standard Johnson V
{\&} R filters plus an H$\alpha $ filter.

For the subsequent analysis, J, H and K data for potential optical
counterparts were obtained from the NASA/IPAC Infrared Science Archive 2MASS
Second Incremental Release Point Source Catalog (PSC) (2000 March)
(http://irsa.ipac.caltech.edu/). B and V data were taken from the Optical
Gravitational Lensing Experiment (OGLE) Stellar Photometric Maps of the
Small Magellanic Cloud  (Udalski et al, 1998.  
ftp://bulge.princeton.edu/ogle/ogle2/maps/smc/).
Accuracy of the zero points of photometry is given as about 0.01 mag, and
astrometry 0.15 arcsec (with possible systematic error up to 0.7 arcsec).

\section{X-ray source location and optical counterpart search}

In order to search for possible counterparts to the X-ray sources, a
set of three selection criteria were defined. These criteria were
partially based upon both the direct observations of other identified
counterparts to BeXBs in the SMC and partially based upon the
possible range of observational characteristics of Be stars in the
SMC.

\begin{figure}
\begin{center}
\psfig{file=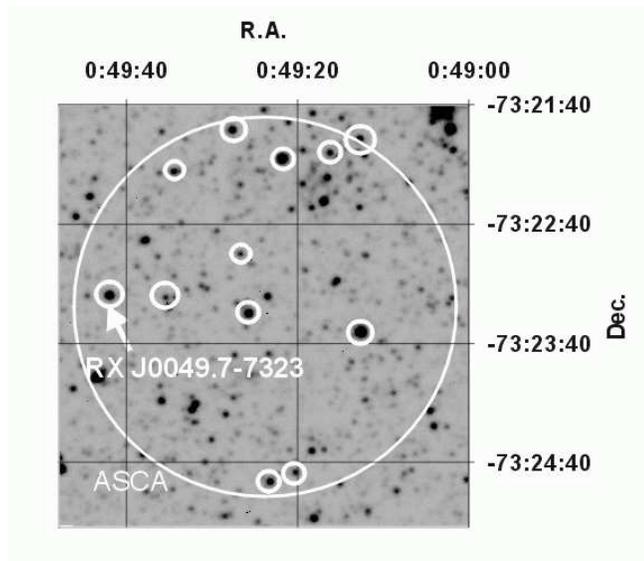,width=8.5cm} \caption{AX J0049.4-7323.
Optical candidates meeting the selection criteria. The
 size of the small white circles around the stars is indicative of the
r-H$\alpha $ values. The large white circle is the ASCA error region. The image
of  the field is taken from the January 2000 V band observation.}
\label{fig:494} \end{center}
\end{figure}

The first criterion, an optical (B-V) colour index, was obtained
by assuming that the stars being searched for were in the spectral
range B0V - B2V and that the extinction to the SMC was somewhere in
the range $0.08<E(B-V)<0.25$. The lower value comes from the work of
Schwering \& Israel (1991) and the upper value from direct observations
of similar Be systems (see, for example, Coe, Haigh and Reig 2000) and
includes a local contribution from the circumstellar disk.  The limits
chosen were $-0.2<(B-V)<0.2$.

The second criterion was simply a cut-off in the V band magnitude at
17.0, again set by assuming the same spectral class range as above
projected to the SMC through any reasonable amount of interstellar and
circumstellar absorption.

The third criterion, an infrared (J-K) colour index was determined
entirely from previous work. Since the circumstellar disk is the
single major contributor to the IR flux, the state of the disk
defines the size of the IR excess. Previously determined values
for (J-K) range from -0.1 up to 0.6, so a limit of $(J-K)<0.7$ was
chosen from these observations.

Optical r-H$\alpha $ colour indices were also computed because a
high relative value would indicate the presence of strong H$\alpha
$. Where the identification of the counterpart is ambiguous, these
are shown in the paper. The R and H$\alpha $ magnitudes were taken
from separate SAAO images and both were uncalibrated. Consequently
the absolute values of this index have no useful meaning and no
selection criterion has been applied.

\subsection{AX J0049.4-7323}

Ueno et al (2000) reported an ASCA observation made during 2000
April 11-17 which revealed coherent pulsations of period 755.5
$\pm $0.6 s from a new source in the Small Magellanic Cloud. This
was designated AX J0049.4-7323 and was located at R.A. 0h 49m
25.2s, Dec. -73$^{o}$ 23' 17" (equinox 2000.0; error radius 1'.5).
See Figure~\ref {fig:494}. The spectrum was characterized by a
flat power-law function with photon index 0.7 and X-ray flux 1.1 x
10$^{ - 12}$ erg cm$^{ - 2}$ s$^{ - 1}$ (0.7-10 keV). They noted
that the possible Be/X-ray binary RX J0049.7-7323 (Haberl and
Sasaki (2000)), was located within the ASCA error region.

\begin{figure}
\begin{center}
\psfig{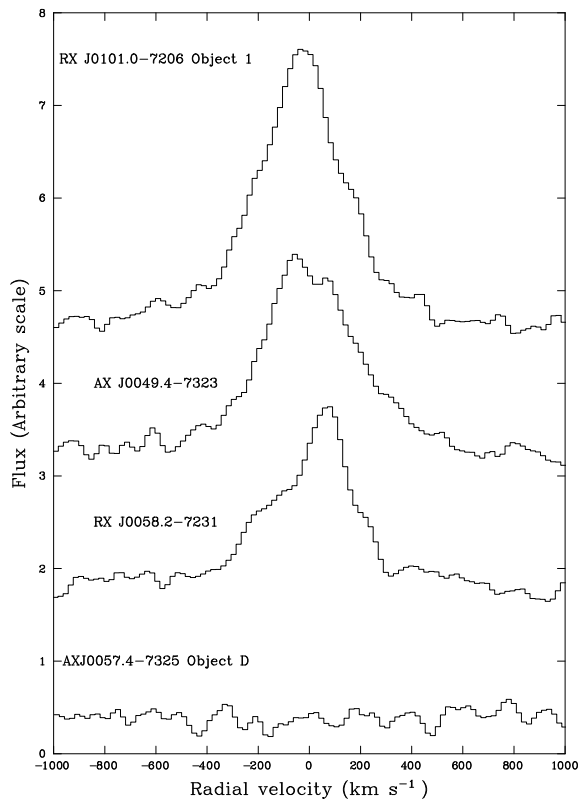} \caption{Offset normalised
profiles of objects showing H$\alpha $ emission. The original data
have been smoothed using a Gaussian of $\sigma =0.5{\AA}$. The
velocity scale has been normalised to the recession velocity of
the SMC. Curves are offset on the flux scale as follows: RX
J0058.2: +0.8, AX J0049.4: +2.2, RX J0101.0: +4. The "wine bottle"
structure of RX J0101.0-7206 is evidence for the presence of a
circumstellar disk observed from close to the axis of rotation.
The double peaked profile of the other two H$\alpha $ lines
indicates the presence of asymmetric discs, viewed from an angle
close to the plane of the disc.} \label{fig:plots}
\end{center} \end{figure}

H$\alpha $ spectroscopy was carried out on 2001 November 7 on the
optical source identified with RX J0049.7-7323. Strong H$\alpha $
emission of equivalent width -23.7$\pm $0.8{\AA} was found
(Figure~\ref {fig:plots}). It is noted that the profile of the
curve exhibits a distinct double peak. This is consistent with
Doppler effects which would be expected from a circumstellar disc
viewed in the plane of rotation. There is also definite V/R
asymmetry between the peaks. This is compelling evidence for the
presence of a Be star and provides further confirmation that this
is the correct identification of this Be/X-ray binary.

\begin{table*}
\begin{tabular}
{|p{15pt}|p{19pt}|p{30pt}|p{23pt}|p{21pt}|p{27pt}|p{34pt}|p{34pt}|p{36pt}|p{34pt}|}
\hline
\multicolumn{3}{|p{64pt}|}{{\bf R.A. (2000)}} &
\multicolumn{3}{|p{71pt}|}{{\bf Dec. (2000)}} &
{\bf J-K}&{\bf V}&
{\bf B-V}&{\bf r-H$\alpha $} \\
\hline 0&49&12.7&-73&23&34.2&0.10&13.86&0.04&-0.25 \\
\hline 0&49&12.7&-73&21&57.1&0.34&15.81&-0.01&0.34 \\
\hline 0&49& 16.2&-73& 22&3.8& 0.20&16.14& -0.01&-0.38 \\
\hline 0& 49&21.8& -73&22& 6.8&0.14& 14.38&0.09& -0.47 \\
\hline 0&49& 23.5&-73& 24&49.0& 0.27&15.62& 0.02&-0.47 \\
\hline 0& 49&25.8& -73&23& 24.2&0.40& 14.95&0.08&-0.39 \\
\hline 0&49& 26.7&-73& 22&54.3& -0.03&16.68& -0.02&-0.73 \\
\hline 0& 49&27.6& -73&21& 52.0&0.20& 15.55&-0.04& -0.28 \\
\hline 0&49& 34.4&-73& 22&12.7& 0.56&15.84& 0.11&-0.62 \\
\hline 0& 49&35.6& -73&23& 15.6&0.36& 16.40&0.08& 0.07 \\
\hline {\bf 0}&{\bf 49}& {\bf 42.0}&{\bf -73}& {\bf 23}&{\bf
14.9}& {\bf 0.26}&{\bf 14.99}& {\bf 0.05}&{\bf 0.01} \\
\hline 0& 49&20.4& -73&24& 45.2&0.26&16.23&-0.17&-0.50 \\
\hline
\end{tabular}
\caption{AX J0049.4-7323. Colour indices of optical candidates meeting the
selection criteria. The source identified with RX J0049.7-7323 is shown in
bold type.} \label{tab1} \end{table*}

Optical photometric values for the whole field within, and immediately
outside, the ASCA error circle were taken from the January 2000 observations
for the B, V, R and H$\alpha $ bands. B and V values were calibrated against
objects selected from the OGLE database. J and K values were taken from the
2MASS Survey. R and H$\alpha $ values could not be calibrated. The relevant
colour indices for all objects meeting the previously defined criteria are
summarised in Table~\ref {tab1}. It is
noted that the source identified with RX J0049.7-7323 was found to have a B-V
value of 0.05$\pm $0.02 and a J-K value of 0.26$\pm $0.12. The V and J
values were 14.99 and 14.50 respectively.These data do not point conclusively
to a single counterpart, and Figure~\ref {fig:494} shows the position of these
objects in relation to the ASCA error circle. The size of the white circles is
proportional to the uncalibrated r-H$\alpha $ values; the bigger the circle the
stronger the H$\alpha $.

The presence of strong H$\alpha $ emission from the optical source identified
with RX J0049.7-7323 and the asymmetric double peak identifies it as a Be star
and points strongly to the conclusion that this is also the optical counterpart
of AX J0049.4-7323. However the possibility remains that one of the other
candidates listed at Table~\ref {tab1} may in fact be the correct identification
because of the large ASCA error circle.

\subsection{AX J0057.4-7325}

Torii et al (2000) reported an ASCA observation made on 2000 April
25-26 which detected coherent pulsations of period 101.45$\pm
$0.07s from a new source in the Small Magellanic Cloud, designated
AX J0057.4-7325 and located at R.A. 0h 57m 27.0s, Dec. -73$^{o}$
25' 31" (equinox 2000.0; error radius 1'). The spectrum was
characterized by a flat power-law function with photon index 0.9
and X-ray flux 2.4 x 10$^{ - 12}$ erg cm$^{ - 2}$ s$^{ - 1}$
(0.7-10 keV). They noted that the ROSAT source, RX J0057.3-7325
(Kahabka et al. 1999) is located within the error region.

\begin{figure}
\begin{center}
\psfig{file=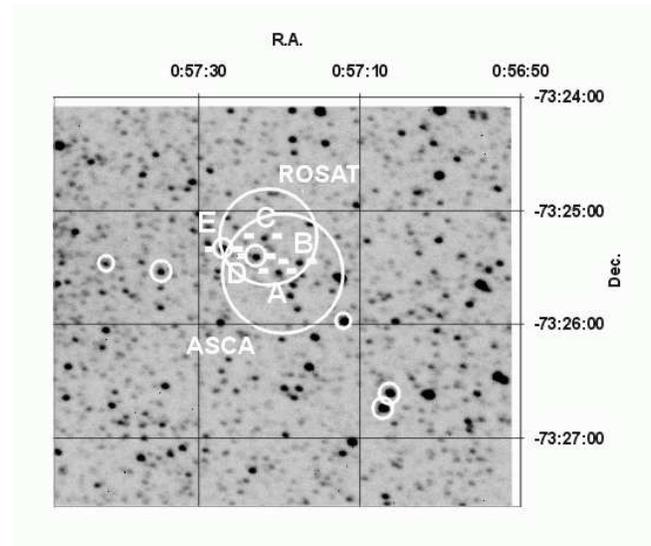,width=8.5cm} \caption{AX J0057.4-7325.
Optical candidates meeting the selection criteria. The
 size of the small white circles around the stars is indicative of the
r-H$\alpha $ values. The larger circles show the ROSAT and ASCA error
regions. The image of  the field is taken from the January 2000 V band
observation.} \label{fig:574} \end{center}
\end{figure}

Four potential counterparts falling within both the ROSAT and ASCA
error circles were identified and designated A, B, C and D
(Figure~\ref {fig:574}). Spectra were taken at SAAO on 2001
November 10. None of these stars, however, showed identifiable
evidence of H$\alpha $ emission above a 2$\sigma $ upper limit of
-2.46 {\AA}.

Using the same IR photometric data as the forgoing section
(regrettably the OGLE B-V data do not cover this field), only one
candidate (D) met the selection criteria. One other object
(labelled E) emerged as a possible candidate within the ROSAT
error circle. See Table ~\ref{tab2}.

\begin{table}
\begin{tabular}
{|c|c|c|c|c|c|c|c|c|}
\hline
\multicolumn{3}{|p{52pt}|}{{\bf R.A. (2000)}} &
\multicolumn{3}{|p{64pt}|}{{\bf Dec. (2000)}} &
{\bf J-K}&
{\bf r-H$\alpha $}&
{\bf ID} \\
\hline
0& 57&27.1& -73& 25& 19.4& 0.08& -0.33& E \\
\hline
\bf 0& \bf 57& \bf 23& \bf -73& \bf 25& \bf 23.4& \bf 0.67& \bf -0.42& \bf D
\\ \hline
\end{tabular}
\caption{AX J0057.4-7325. J-K and r-H$\alpha $ colour indices of two
 candidates meeting the selection criteria. Object D is shown in bold type.}
\label{tab2}
\end{table}

Based on the photometric data, Object D therefore emerges as the most likely
candidate for this source, although the absence of any detectable H$\alpha $
emission in the spectrum taken in November 2001 indicates
that no strong conclusion can be drawn. It is possible that time variability may
account for the fact that H$\alpha $ was not observed in emission.

\subsection{RX J0058.2-7231}

\begin{figure}
\begin{center}
\psfig{file=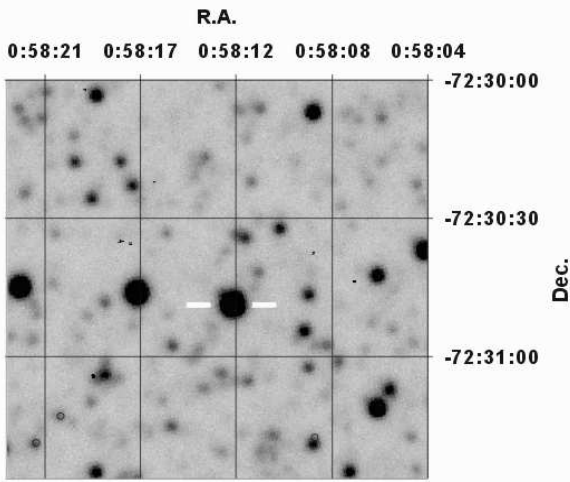,width=8.5cm} \caption{RX J0058.2-7231.
The likely optical counterpart is indicated by the white lines.
The image of the field is taken from the January 2000 V band
observation.} \label{fig:582}
\end{center}
\end{figure}

Schmidtke et al (1998) reported the detection of this very weak X-ray source,
with a count rate slightly below the 3$\sigma $ level, by
the ROSAT High-Resolution Imager (HRI). They noted that it was located only
5" from a 15$^{th}$ magnitude Be star which was likely to be the optical
counterpart (Figure~\ref{fig:582}).

Spectroscopy carried out on this star on 2001 November 9 revealed
H$\alpha $ emission of equivalent width -14.57$\pm $0.52{\AA}
(Figure~\ref{fig:plots}). There is a very marked V/R asymmetry in
the line profile which can again be attributed to a global
one-armed oscillation providing evidence for the presence of a
circumstellar disc viewed in the plane of the disc, and thus
evidence for the detection of a Be star. The B-V colour index was
measured at 0.06$\pm $0.02 and the J-K index at 0.39$\pm $0.10.
The V and J values are 14.90 and 14.53 respectively. These
values meet the previously defined criteria, providing strong
confirmation that this is indeed the optical counterpart.
Furthermore this is the only star of $V<17th$ magnitude within the
uncertainty of 5" give by Schmidtke et al.

\subsection{RX J0101.0-7206}

The X-ray transient RX J0101.0-7206 was discovered in the course of ROSAT
observations of the SMC in October 1990 (Kahabka and Pietsch 1996) at a
luminosity of 1.3 x 10$^{36}$ erg s$^{ - 1}$. The source was seen in
outburst for 22 hours but half a year later was no longer detectable at a
2$\sigma $ upper limit luminosity of 4.6 x 10$^{34}$ erg s$^{ - 1}$.

\begin{figure}
\begin{center}
\psfig{file=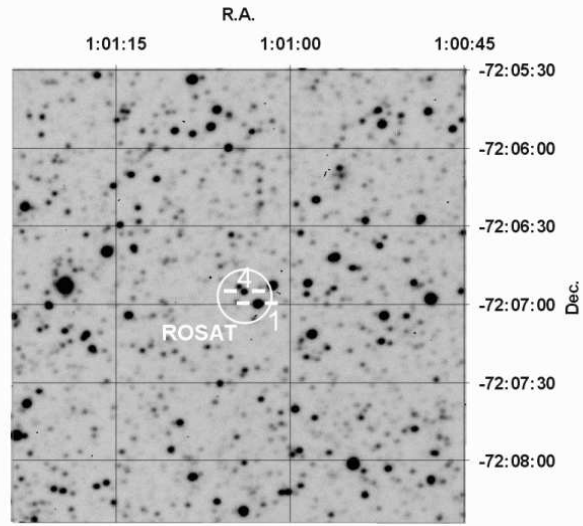,width=8.5cm} \caption{RX J0101.0-7206.
Object 1 is the likely optical counterpart. The circle
 indicates the ROSAT error region. The image of the field is taken from
 the January 2000 V band observation.}
\label{fig:010}
\end{center}
\end{figure}

H$\alpha $ spectra of the only two clearly visible stars in the
error circle were taken on 2001 November 9. These are numbered 1
and 4 on the V band image
 (Figure~\ref{fig:010}). Object 1 was found to emit H$\alpha $ of equivalent
width -54.6$\pm $1.3{\AA} (Figure~\ref{fig:plots}). It is noted
that the profile is single peaked with a "wine bottle" shape. This is evidence
for the detection of a Be star observed from an angle close to the axis of
rotation (Slettebak et al 1992). H$\alpha $ emission above a 2$\sigma $ upper
limit of $-3.1{\AA} $ could not be detected from Object 4, which also failed to
meet the photometric criteria.

Object 1 also met the photometric criteria applied to the objects in
previous sections, showing a B-V colour index of -0.04$\pm $0.02 and a J-K index
of 0.05$\pm $0.26. The V and J values were 15.74 and 15.55
respectively. Furthermore, this object exhibited the highest r-H$\alpha $ value
of all qualifying objects in the image. We therefore conclude that Object 1 is
the optical counterpart.

\section{Discussion and Conclusions}

We have confirmed the identification of RX J0049.7-7323 with its
presumed optical counterpart. The presence of strong H$\alpha$
emission and the clear double peaked spectrum from this optical source
points strongly to the conclusion that it is also the optical
counterpart of AX J0049.4-7323, however the possibility remains that
one of the other candidates identified may be the source associated
with the latter. From its (B-V) colour we can infer a spectral type in
the range B1-3V assuming the higher extinction limit of $E(B-V)=0.25$
seen in other similar systems in the SMC. However, because of the
uncertain contribution to the extinction by the circumstellar disk,
the only reliable way of determining the spectral class of these
objects is by obtaining detailed blue spectra.

The absence of any H$\alpha $ emission from any of the candidates for AX
J0057.4-7325 prevents a confident identification, although Object D in
Figure~\ref{fig:574} emerges as the most likely counterpart.

Similarly, we conclude that the strong H$\alpha $ emission and marked V/R
asymmetry of the line profile from the suggested optical counterpart
to RX J0058.2-7321 are confirmation that this is a Be star and hence
probably the correct identification. From its (B-V) colour index we
can infer that it is a B2-3V type star.

Finally we conclude that Object 1 in Figure~\ref{fig:010} is the
optical counterpart to RX J0101.1-7206. It has a slightly higher (B-V)
colour index suggesting a slightly later spectral type, maybe
B3-4V. However, we note that the H$\alpha$ EW was much larger for this
source than any of the others and hence the larger circumstellar disk
implied could well be further modifying the colour index.

Future blue spectra of these systems from a larger telescope
should help establish the correct spectral classification for all
these objects and finally resolve the remaining uncertainties.

\section{Acknowledgments}

We are grateful to the staff of SAAO for their support during our
observations.This research has made use of the NASA/ IPAC Infrared Science
Archive, which is operated by the Jet Propulsion Laboratory, California
Institute of Technology, under contract with the National Aeronautics and Space
Administration. It has also made use of the Optical Gravitational Lensing
 Experiment
(OGLE) Stellar Photometric Maps of the Small Magellanic Cloud, through the
Princeton University site. Finally, we have also drawn on the observations
 carried
out by J B Stevens at the SAAO in 1996.

\end{document}